\documentclass[%
 reprint,
 amsmath,amssymb,
 aps,
]{revtex4-2}

\usepackage{graphicx}% Include figure files
\usepackage{hyperref}
\usepackage{dcolumn}% Align table columns on decimal point
\usepackage{bm}% bold math
\usepackage{chngcntr}
\begin{document}

\preprint{APS/123-QED}

\title{Fabric-Based Star Soft Robotic Gripper}

\author{Ignacio Andrade-Silva}
 \email{ignacioandradesilva@gmail.com}

\author{Joel Marthelot}%
 \email{joel.marthelot@univ-amu.fr}
\affiliation{%
 Aix-Marseille Univ, CNRS, IUSTI, 13013 Marseille, France
}%

\date{\today}

\begin{abstract}
Soft pneumatic gripping strategies are often based on pressurized actuation of structures made of soft elastomeric materials, which limits designs in terms of size, weight, achievable forces, and ease of fabrication. In contrast, fabric-based inflatable structures offer high stiffness-to-weight ratio solutions for soft robotics, but their actuation has been little explored. Herein, a new class of pneumatic soft grippers is presented that exploits the in-plane overcurvature effect of inextensible fabric flat balloons upon inflation. A star-shaped gripper contracts radially under pressure producing a gripping force on the object whose intensity can be modulated by the pressure input. First, the kinematics and mechanics of a single V-shaped actuator are studied through experiments, finite element simulations, and analytical models. Then, these results are leveraged to predict the mechanical response of the entire star, optimize its geometry, and maximize contraction and stiffness. It is shown that the gripping performance can be improved by stacking several stars with silicon-coated corners. It is expected that the flexibility, robustness, scalability, and ease of fabrication of this methodology will lead to a new generation of lighter and larger actuators capable of developing higher forces and moving delicate and irregularly shaped objects while maintaining reasonable complexity.
\end{abstract}

\maketitle

%\tableofcontents

\section{Introduction}

Designing robots made out of soft elastomeric materials has opened up new possibilities for performing tasks where conventional robots fail, such as for grasping and manipulating delicate or irregular objects\,\cite{rus2015design,brown2010universal,ilievski2011soft,jones2021bubble}, complex locomotion and navigation\, \cite{shepherd2011multigait,hawkes2017soft}, or assisting in complex biological functions, from wearable applications\, \cite{polygerinos2015soft,yap2017fully,thalman2018novel,sanchez2021textile,o2021unfolding, rajappan2022logic} to heart muscle contraction\,\cite{roche2014bioinspired}. Unlike their rigid counterparts, soft robots can safely interact with humans and require neither environment perception systems nor precise position detection of the objects they interact with. Fluid-driven actuators\, \cite{acome2018hydraulically}, e.g. silicone-bodied pneumatic actuators powered by pressurized voids\,\cite{martinez2013robotic, overvelde2015amplifying, gorissen2017elastic, siefert2019bio}, are widely used due to their simple and rapid actuation, easy operation, and scalable implementation. The kinematics of these actuators is directly programmed into their anisotropic structure which stretches differentially under pressure and induces their contraction, expansion, torsion, or flexion. However, these deformation modes that involve large material strains require the use of rubber-like materials (e.g., elastomers). Because of their low stiffness, they generally need thick walls, which results in large unpressurized volumes and limits their designs to small scales (a few centimeters), as larger structures tend to collapse under their own weight. 

To use stiffer materials that can be actuated at larger scales, other strategies have been proposed, exploiting the bending deformation of thin structures that remain within the inextensible limit of the material. The actuation process is embedded within networks of cuts or folds to gently grip objects\,\cite{yang2021grasping, hong2022boundary}, morph\,\cite{andrade2019foldable,yu2021cutting}, or deploy\,\cite{filipov2015origami, li2017fluid, melancon2021multistable}. 
Fabric-based inflatable actuators are particularly attractive because they are generally simple to manufacture, foldable, lightweight, require low operating pressure, can be used at larger scale\,\cite{amase2015mechanism, siefert2020programming, panetta2021computational, usevitch2020untethered,jones2023soft} and develop higher forces\,\cite{li2019vacuum, nishioka2017development} than their elastomeric counterparts. 

In this work, we present a simple strategy for designing a soft gripper based on the deformation of inextensible fabric star-shaped inflatables that contracts radially in a plane under pressure. The gripper consists of a network of flat tubes connected by V-shaped actuators (V-actuator) that act as hinge-like mechanisms. Using experiments, finite element simulations, and theoretical modeling, we rationalize the actuation process and model the mechanical response of a V-actuator as a function of its geometry, applied pressure, and mechanical properties of the fabric. We leverage this quantitative knowledge to program the deformation of the network of tubes and calculate the optimal star shape maximizing the radial contraction and the force applied to the target object. Unlike their rubber counterparts, our inflatable soft grippers can fold flat when not pressurized and have a high stiffness-to-weight ratio, making it easy to extend their design to large scales.

\begin{figure*}[t]%[\sidecaptionrelwidth]
\centering
\includegraphics[width=16cm]{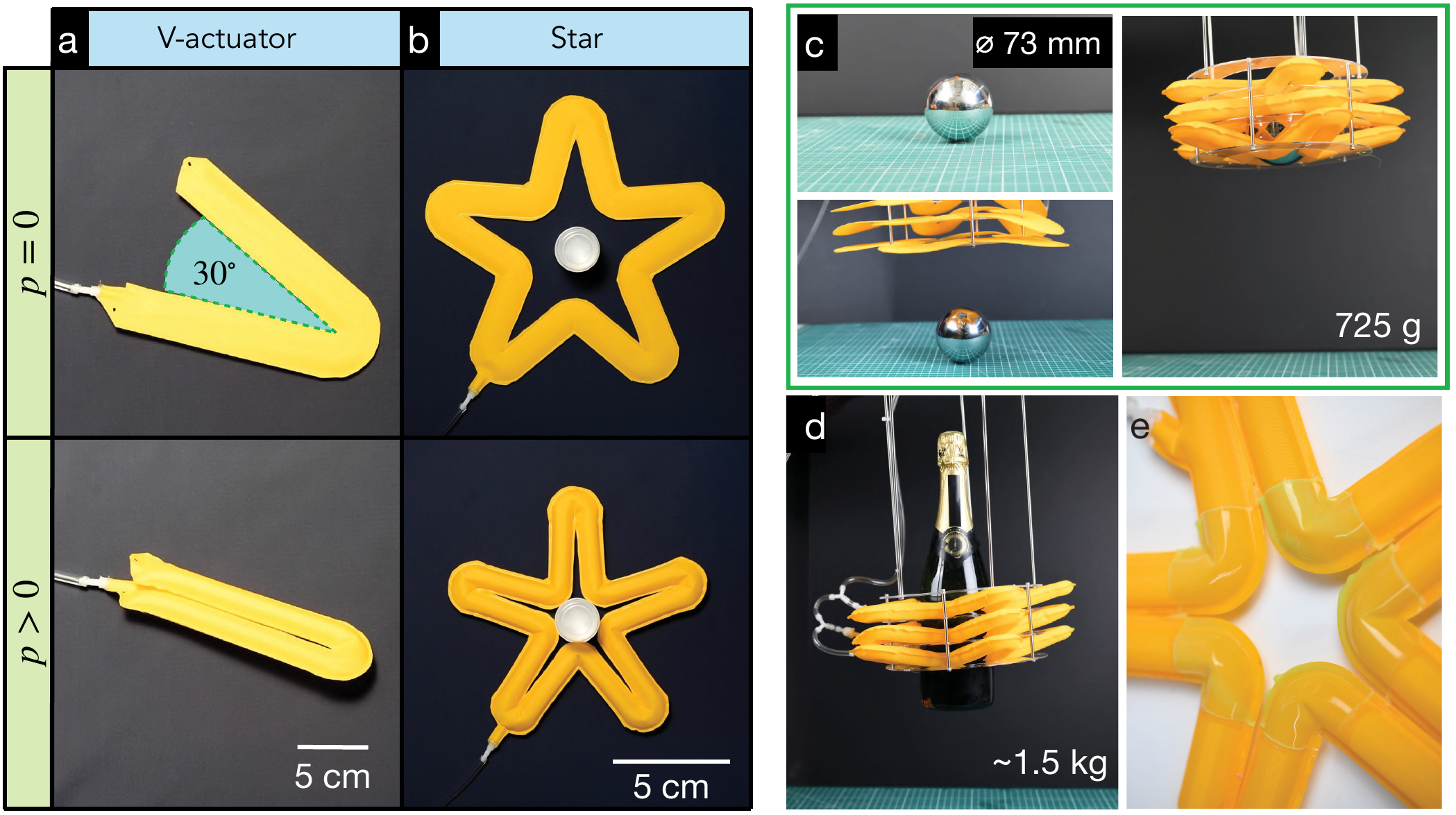}
\caption{Overcurvature-based fabric star gripper. (a) A V-actuator closes on itself upon inflation. (b) A 5-pointed star composed of 10 alternated V-actuators contracts radially in the plane. (c-d) Gripper composed of three 5-pointed stars stacked vertically in an acrylic frame. The frame allows the easy positioning of the deflated structure. The device is shown gripping various objects: (c) a p\'etanque ball and (d) a champagne bottle. (e) Coating of the inner corners of the star with a silicone based elastomer.
}\label{fig:star}
\end{figure*}

%\section*{Results}

Figure\,\ref{fig:star} illustrates our approach. Our model consists of two nylon fabric sheets sealed together to form a network of inflatable tubes. The fabric sheet of thickness $t=300\,\mu$m, Young's modulus $E=128$\,MPa, and Poisson ratio $\nu=0.39$ is characterized by a stretching modulus $Y=Et$, and a bending modulus $B=Et^3/[12(1-\nu^2)]$. The two sheets are heat-sealed along the desired path using a soldering iron mounted on a two-axis translation stage (Experimental Section). A V-actuator consists of two rectangular parts connected by an annulus sector forming an angle $\phi_0$. Upon inflation, a V-actuator with an initial acute angle of $\phi_0=30^\circ$ in the flat configuration closes completely on itself to a target angle $\phi_t<0^\circ$, acting as a hinge-like mechanism (Fig.\,\ref{fig:star}a). We take advantage of this closing mechanism to build networks of tubes connected by V-actuators. By concatenating a series of actuators in alternate directions, we build star-shaped closed loops. During inflation, the tube network contracts radially in the plane (Fig.\,\ref{fig:star}b). By adding an acrylic frame to provide rigidity to the unpressurized structure, we show that this overcurvature effect can be harnessed to design a soft gripper able to grip objects of different shapes and weights. The actuation process can be easily stacked to increase the gripping force and conform to objects of different geometries, from metal spheres to bottles and vases (Fig.\,\ref{fig:star}c,d and Movie S1).
Friction with the object is increased by coating the inner corners of the star with a silicone-based elastomer (Fig.\,\ref{fig:star}e). The prototype achieves remarkably high forces, being able to grip objects weighing several kilograms. We first model the kinematic and mechanical response of an individual V-actuator before extending our approach to networks.

\section{Response of a single V-actuator}
\subsection{Kinematic response of a V-actuator}
\begin{figure*}%[tbhp]
\centering
\includegraphics[width=16cm]{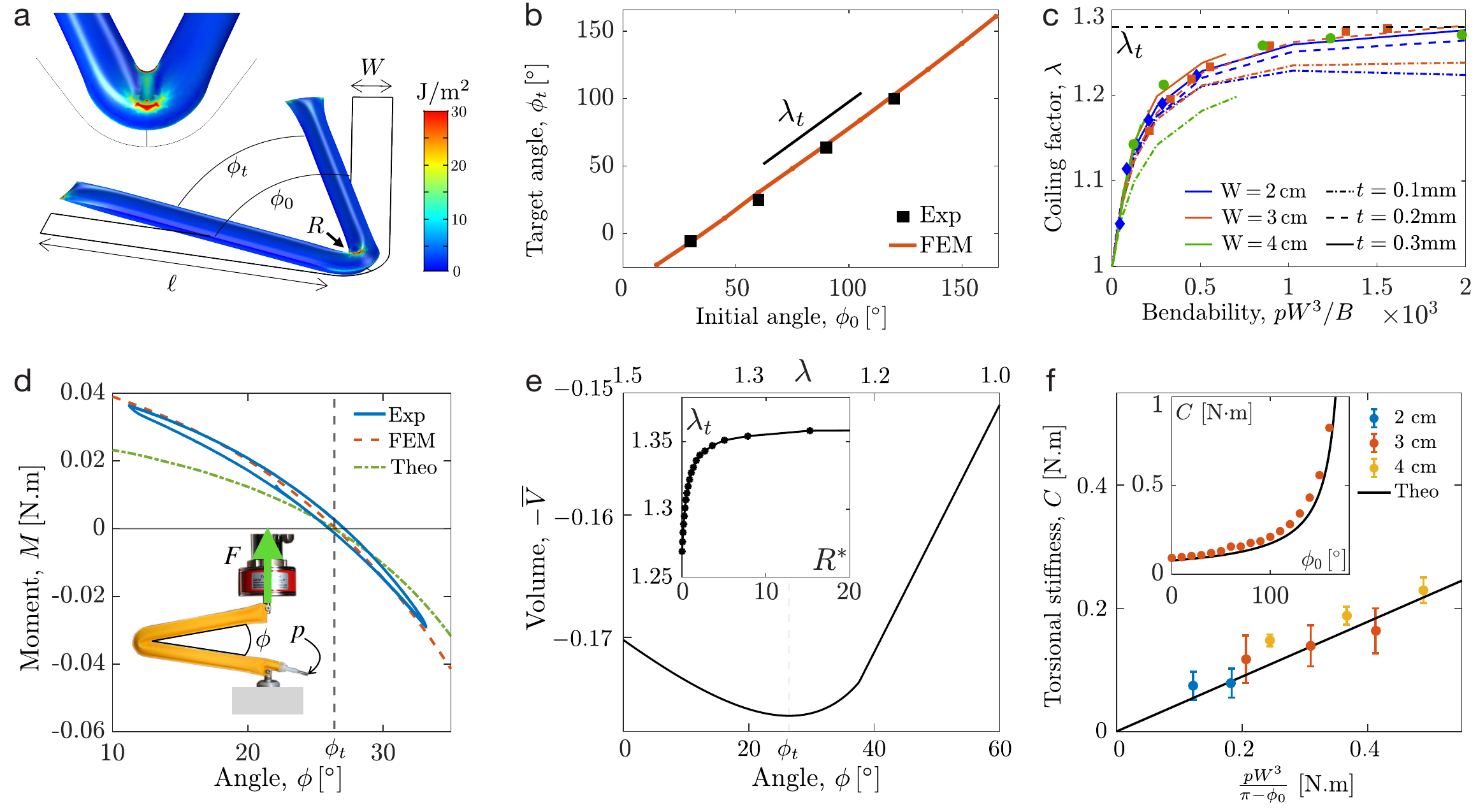}
\caption{Kinematic and mechanical response of a single V-actuator. (a) FEM simulation of a single V-actuator showing overcurvature ($W=2$ cm, $R = W/10$, $\ell=15$ cm, $\phi_0 = 75^{\circ}$ and $p=14$kPa). The inset shows a zoomed top view of the actuator where stress focusing can be seen. Colors code the strain energy density. (b) Target angle $\phi_t$ shows a linear relation with the initial angle $\phi_0$ ($W = 2$ cm). The slope corresponds to the coiling factor. (c) Actual coiling factor $\lambda$ as a function of the pressure for constant $R^*=R/W=0.1$. Lines correspond to numerical simulations while dots are experimental data for $t=0.3$ mm. The black dashed line is the theoretical prediction $\lambda_t$ valid at high pressure.
(d) Moment $M(\phi)$ as function of its angle $\phi$ measured from the experiment (blue) and extracted from FEM simulations (red, dashed), with $W=2$ cm, $R^*=0.1$, $\phi_0=60$º and $p=13.8$ kPa. Theoretical prediction of volume change contribution on the moment neglecting strain variation is plotted as a green line.
Inset: Snapshot of the experimental setup.
(e) Theoretical volume change of the hinge section as function of $\phi$ ($\phi_0 = 60$º) showing the pressure energy as an asymmetric well potential. The minimum of pressure energy occurs at $\phi_t$. Inset: Theoretical target coiling factor $\lambda_t$ versus $R^*$.
(f) Hinge stiffness extracted from FEM for three different widths ($W=2,\, 3, \,4 $ cm, $\phi_0 =30$º). The solid black line corresponds to the theoretical prediction.
Inset: FEM extracted stiffness versus $\phi_0$ ($W=3$cm, $p=20$ kPa).}
\label{fig:overcurvature}
\end{figure*}

We first proceed to the experimental and numerical characterization of the kinematics of a single V-actuator, consisting of two rectangular arms of width $W$ and length $\ell$ connected by an annulus sector with an inner radius of curvature $R$. Figure\,\ref{fig:overcurvature}a shows the deformed shape of an actuator predicted by finite element methods (Experimental Section, and Finite Element Simulations, Supplementary Material). During inflation, the angle between the two arms decreases from the initial angle $\phi_0$ in the flat configuration to a target angle $\phi_t$ which is obtained at high pressures (Movie S2).
The sheet deforms strongly in the vicinity of the edge (see Fig.\,\ref{fig:overcurvature}a Inset), leading to a geometric localization of the deformation close to the bend reminiscent of the Brazier instability observed in thin-walled cylindrical tubes or the curvature condensation observed in thin elastic sheets\,\cite{brazier1927flexure,das2007curvature, qiu2022bending}.

In Fig.\,\ref{fig:overcurvature}b, we plot the target angle $\phi_t$ of tubes made with different $\phi_0$ and the same width $W=2$\,cm, length $\ell=15$\,cm and radius of curvature $R=2$\,mm. Finite elements simulations plotted in red line are in excellent agreement with experimental observations. We observe a linear relationship in which angular contraction is zero at $\phi_0=\pi$ and maximum at $\phi_0=0$. We define the actual coiling factor of the structure $\lambda(\phi)=(\pi-\phi)/(\pi-\phi_0)$, where $\phi$ is the actual angle at any stage of inflation. 

Figure\,\ref{fig:overcurvature}c shows the actual coiling factor $\lambda$ as a function of the pressure $p$ measured from both experiments and simulations, showing that $\lambda$ increases with the operating pressures and saturates at the target coiling factor $\lambda_t=\lambda(\phi_t)$. The data for different tube widths and membrane thicknesses (Fig.\,\ref{fig:overcurvature}c) fall on a master curve showing that the coiling process is dominated by the bendability of the membrane $\epsilon^{-1}=pW^3/B$. This plot also predicts the operating pressure that must be applied to the actuator to achieve the maximum change in angle which is of the order of $10^3B/W^3$. 

Upon inflation, the system minimizes the total energy $U_T = U_{s} - pV$, where $U_{s}$ is the strain energy, $p$ the fluid pressure and $V$ is the total volume enclosed by the system. In the case of an infinitely thin membrane, the total energy is dominated by $-pV$, whose minimization is equivalent to the maximization of the volume\,\cite{siefert2019programming, siefert2020geometry}.
The typical operating pressure of the order of 10\,kPa is too small to induce stretching in the fabric ($p<Y/W$) so that the membrane can be considered quasi-inextensible, therefore the resulting shape is a consequence of volume maximization under inextensibility constraints\,\cite{davidovitch2011prototypical, paulsen2015optimal}. 
The shape of the curved hinge can be approximated to that of a sector of an axisymmetric ring whose cross-sectional profile is obtained by a Lagrangian model maximizing the volume under inextensibility constraints (Inflation of an annular sector, Supplementary Material). This optimization problem leads to an optimal coiling factor $\lambda_t(R^*)$ that only depends on $R^*=R/W$. In our case, the optimal coiling $\lambda_t(R^*=0.1)=1.28$ plotted as a dashed line in Fig.\,\ref{fig:overcurvature}c captures the asymptotic behavior of the actuator at large pressure for all geometries. 

\subsection{Mechanical response of a V-actuator}
We now study the mechanical hinge-like response of the V-actuator around its target angle $\phi_t$. The actuator is attached to a universal testing machine with pin-joint conditions (see Experimental Section and inset in Fig.\,\ref{fig:overcurvature}d), while the pressure imposed in the tube is kept constant. We measure the vertical force $F$ as the V-actuator opens or closes and compute the hinge moment defined as $M_h=F\,\ell \,\cos(\phi/2)$.
We couple this force measurement with the direct observation of the actuator deformation with a camera to compute the moment applied by the structure $M_h$ as a function of the imposed angle $\phi$ (Fig.\,\ref{fig:overcurvature}d). 

We observe that the mechanical response of the hinge presents a small hysteresis and that the moment is highly non-linear around the target angle. The hysteresis arises from fabric mechanical response and the fact that the closed volume changes as the pressure valve restricts airflow.%comes from the fact that the enclosed volume is changing while the pressure valve restricts air outflow.
The hinge mechanism is softer when closed ($\phi>\phi_t$) and stiffer when opened ($\phi<\phi_t$). The experiments are in excellent agreement with numerical simulations plotted as a red line. 
%The force was set to zero after mounting the actuator to avoid the weight of the structure affects the force measurements. The air supply tube was choosing to be as small and soft as possible and we check that its effect on the force measurement was minimal.

To build a fundamental understanding of the mechanical response of the pressurized actuator, we first decompose the moment the sum of the volume variation and the strain energy variation with respect to $\phi$:
\begin{equation}
    M_T=-\frac{d U_s}{d \phi}+p\frac{d V}{d \phi}.
    \label{eq:M}
\end{equation}
While the variation of the strain energy depends on the localization of the strain energy in the hinge and is difficult to predict analytically, our analytical model allows us to predict the volume change as a function of the angle as we show next. We checked numerically that the mechanical response is dominated by the volume variation, as expected in the thin membrane limit.

When the angle of the actuator changes, the main volume variation comes from the hinge and not from the arms. Following the same theoretical framework developed to model the actuator shape at high pressure, we predict the change in volume $V$ with the imposed angle $\phi$. Figure~\ref{fig:overcurvature}e shows $-\bar{V}(\phi)$ for $R^*=0.1$ and $\phi_0=60^\circ$, where $\bar{V}=V/[W^3(\pi-\phi_0)]$ is the dimensionless volume of the hinge section (Inflation of an Annular Sector, Supplementary Material). The minimum of the curve corresponds to the target angle $\phi_t$ obtained when no force is applied to the actuator. The curve has a parabolic-like shape showing a distinctive asymmetry with respect to its minimum. In the membrane limit, this asymmetric potential is directly responsible of the asymmetrical mechanics of the V-actuator, which results in a stiffer response when the V-actuator is open and a softer response when the V-actuator is closed 

More generally, the variation of the volume can be expressed as a function of the coiling factor $\lambda$ for any given initial angle $\phi_0$ as shown on the upper axis of Fig.~\ref{fig:overcurvature}e. The target coiling factor, $\lambda_t$, can be adjusted slightly by changing the slenderness of the actuator as shown in inset. From Eq.\,(\ref{eq:M}), we deduce the volume-variation contribution to the moment which is plotted as a dashed green line in Fig.\,\ref{fig:overcurvature}d. 

The volume variation with respect to $\phi$ also gives insights into the mechanical stiffness of the V-actuator defined by $C = - \left. d M_T /d \phi \right|_{\phi_t}.$
The volume change contribution to the torsional stiffness of the actuators is:
\begin{equation}
    C \approx \left. -p \frac{d^2 V}{d\phi^2} \right|_{\phi_t} =- \left. \frac{p W^3}{(\pi-\phi_0)} \frac{d^2 \bar{V}}{d\lambda^2} \right|_{\lambda_t}.
    \label{eq:C}
\end{equation}
where the term $d^2 \bar{V}/d\lambda^2 \vert_{\lambda_t}$ depends only on $R^*$. From our theoretical model, we numerically compute this term for different values of $R^*$, finding that it is well fitted by $d^2 \bar{V}/d\lambda^2 \vert_{\lambda_t}=-(0.4659 R^* + 0.3986)$ (Fig. S2C, Supplementary Material). 

The theoretical prediction of Eq.\,(\ref{eq:C}) is plotted in Fig.\,~\ref{fig:overcurvature}f for a given $R^*$ and compared to the stiffness extracted from FEM for three different tube widths. The errorbars correspond to the two different ways of computing the moment numerically based on the hinge description $M_h$ (valid when $W\ll L$) or the generalized definition $M_T$ (see Eq.\,(\ref{eq:M})). In Inset, we plot the stiffness as function of $\phi_0$ which is well captured by Eq.\,(\ref{eq:C}). The torsional stiffness increases for larger initial angles and diverges at $\phi_0 = \pi$. In the following, we approximate the mechanical behavior of the actuator around the target angle to a torsional hinge with linear stiffness: $M(\phi)=-C(\phi-\phi_t)$.

\begin{figure}%[tbhp]
\centering
 \includegraphics[width=\columnwidth]{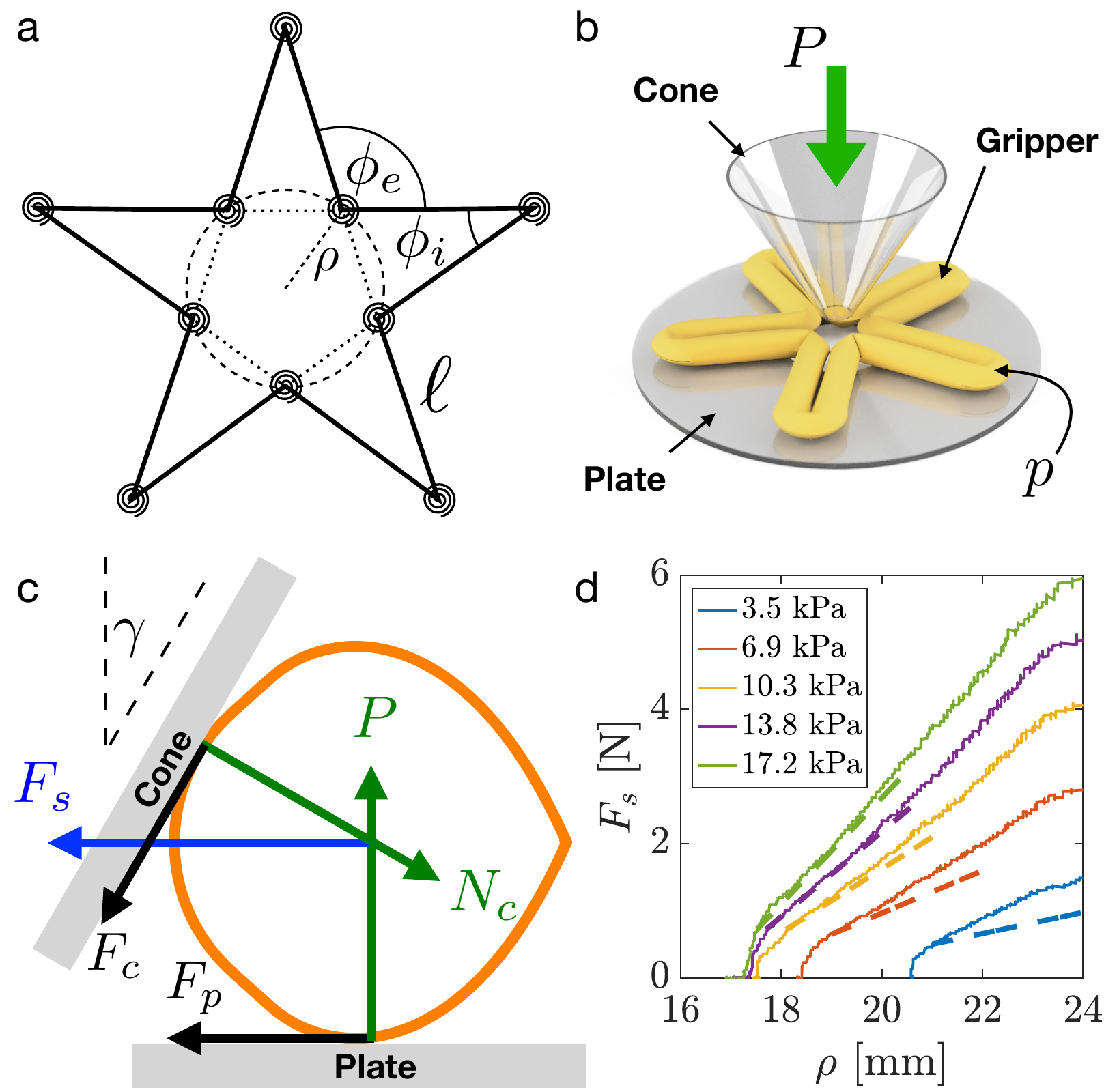}
\caption{Radial contractile force of the gripper. (a) Schematic of an idealized star without width. (b) Setup composed of glass cone pushed in the center of the star which is supported by a glass disk. (c) Force diagram of the cross section of the gripper in contact with the cone. (d) Measured radial force versus inner radius of the gripper (solid lines) and theoretical prediction of the linear response (dashed lines).}
\label{fig:setup}
\end{figure}

\section{Star soft gripper}
\subsection{Kinematics of the star gripper} We now take advantage of the actuator closing mechanism to build a soft gripper based on several V-actuators connected in alternating directions in a closed loop. The geometry of our soft gripper is a $n$-pointed star that is constructed from a $n$-sided regular polygon and isosceles triangles whose bases coincide with the sides of the polygon (Fig.~\ref{fig:setup}a). The triangles have arm lengths $\ell$ and internal angles $\phi_{i0}$ at their apex. The external angle between the arms of two adjacent triangles is given by $\phi_{e0}=\phi_{i0} + 2\pi/n$. Since the arms of the star always remain straight,
we can write:
\begin{equation}
    \phi_e=\phi_i +2\pi/n. \label{eq:geometric_relation}
\end{equation}
where $\phi_{i}$ and $\phi_{e}$ refer to the internal and external angles of the star at any stage of inflation.

When width of the tube is negligible ($W \ll \ell$), the actual radius of the circle enclosed by the star is $\rho(\phi)=\ell \sin{(\phi_i/2)} / \sin{(\pi/n)}$. Notice that in this limit, when $\phi_i \rightarrow 0 $, the star closes completely.

We model our soft gripper as a set of rigid bars and linear torsional hinges (Fig.~\ref{fig:setup}a). The mechanical energy of the system is therefore $E_{star}=n/2 [C_i (\phi_i-\phi_{it})^2 + C_e (\phi_e-\phi_{et})^2]$, where $\phi_{it}$ (resp. $\phi_{et}$) are the internal (resp. external) target angles of the star and $C_i$ and $C_e$ are the corresponding torsional stiffness. Note that $\phi_{it}$ and $\phi_{et}$ do not necessarily satisfy Eq.~(\ref{eq:geometric_relation}).

The minimisation of the energy with respect to $\phi_i$ allows us to predict the internal angle of equilibrium with respect to $\lambda$: 
\begin{equation}
    \phi_i^{eq}=\phi_{it}+\frac{2 \pi}{n}(\lambda_t -1)\left(\frac{C_e}{C_i+C_e}\right).
    \label{phii}
\end{equation}

For small values of $n$, the interaction between the internal and external torsional stiffnesses restricts the complete closure of the star. This interaction depends on the factor $C_e/(C_i+C_e) \approx 1/[2(1-\pi/(\pi-\phi_{0i})/n)]$, whose value tends asymptotically  to 1/2 for large values of $n$. In the zero-width limit, the equilibrium radius is given by $\rho_{eq}=\rho(\phi^{eq}_i)$.
%\sim 1/2 (1 +\pi/(\pi-\phi_{i0})/n )+\mathcal{O}(1/n^2)$

\subsection{Grip strength of the soft gripper}  
When the gripper catches an object, it applies an inwards radial force $F_s$ if the size of the object is larger than the equilibrium radius of the star $\rho_{eq}$. To measure the grip strength as a function of the object diameter, we design an experiment where a rigid cone is pushed through the center of the pressurized star (Fig.~\ref{fig:setup}b) while the star is supported on a plate with a hole in its center. The bottom plate is fixed while the cone is attached to the load cell of a universal testing machine. The cone moves downwards with a constant imposed velocity so that the star opens radially while the force applied to the cone is measured. Both the cone and the plate are made of glass, which ensures a smooth sliding of the fabric, with a well defined coefficient of kinetic friction. The hole in the center is large enough to allow the star to open a few centimeters in diameter before the cone contacts the plate.

Figure~\ref{fig:setup}c shows the cross section of the corner of the star in contact with the cone and the plate. The forces acting on this cross section are: the radial contraction force $F_s$ of the gripper; two normal forces $N_c$ and $P$ acting at the contacts with the cone and plate; two friction forces $F_c$ and $F_p$ at the contacts with the cone and the plate.
A force balance gives the radial force $F_{s}$ as a function of the vertical load $P$ which is measured by the load cell:
\begin{equation}
    F_s = \left(\mu + \frac{1- \mu \tan{\gamma}}{\mu +\tan{\gamma}} \right) P,
\end{equation}
where $\gamma=30^\circ$ is half the angle of the cone and $\mu=0.1$ is the kinetic friction coefficient.

The experimental radial force is plotted as a function of the inner radius of the gripper in Fig.~\ref{fig:setup}d for different operating pressure. The contact point between the star and the cone depends on the equilibrium radius of the star. We predict the radial contractile force by differentiating the mechanical energy of the system with respect to the radius. At linear order, we find that:
\begin{equation}
    F_s=\frac{dE_{star}}{d\rho}\approx\frac{4n(C_i+C_e)\sin^2{(\phi_{i0}/2)}}{\rho^2_0 \, \cos^2(\phi_i^{eq}/2)} \Delta \rho.
    \label{Fs}
\end{equation}
where $\Delta \rho = \rho - \rho_{eq}$. 
By deriving the values of $C_i$ and $C_e$ from Eq.\,(\ref{eq:C}), Eq.\,(\ref{Fs}) give us a prediction for the radial stiffness.
We observe a slight offset in the force at $\rho = \rho_{eq}$ due to the change in volume associated with the local deformation at the cone contact. This additional force should scale with the pressure multiplied by the area of the contact zone, which depends on the geometry of the object. 
The linear stiffness prediction works best for larger pressures, consistent with the fact that in this limit, the energy term $-pV$ is dominant over the material's strain energy.

We define the radial stiffness of the gripper as $K=F_s/\Delta \rho$. Notice that, for large values of $n$, this stiffness scales as $K \sim ~n\, p\, W^3/\rho_0^2$ from Eq.\,(\ref{eq:C}) and Eq.\,(\ref{Fs}). Equation\,(\ref{Fs}) shows that at fixed $\rho_0$ and $\phi_i$, a larger number of points $n$ implies a smaller arm length $\ell$, resulting in greater stiffness. 

\begin{figure*}%[tbhp]
\centering
\includegraphics[width=14cm]{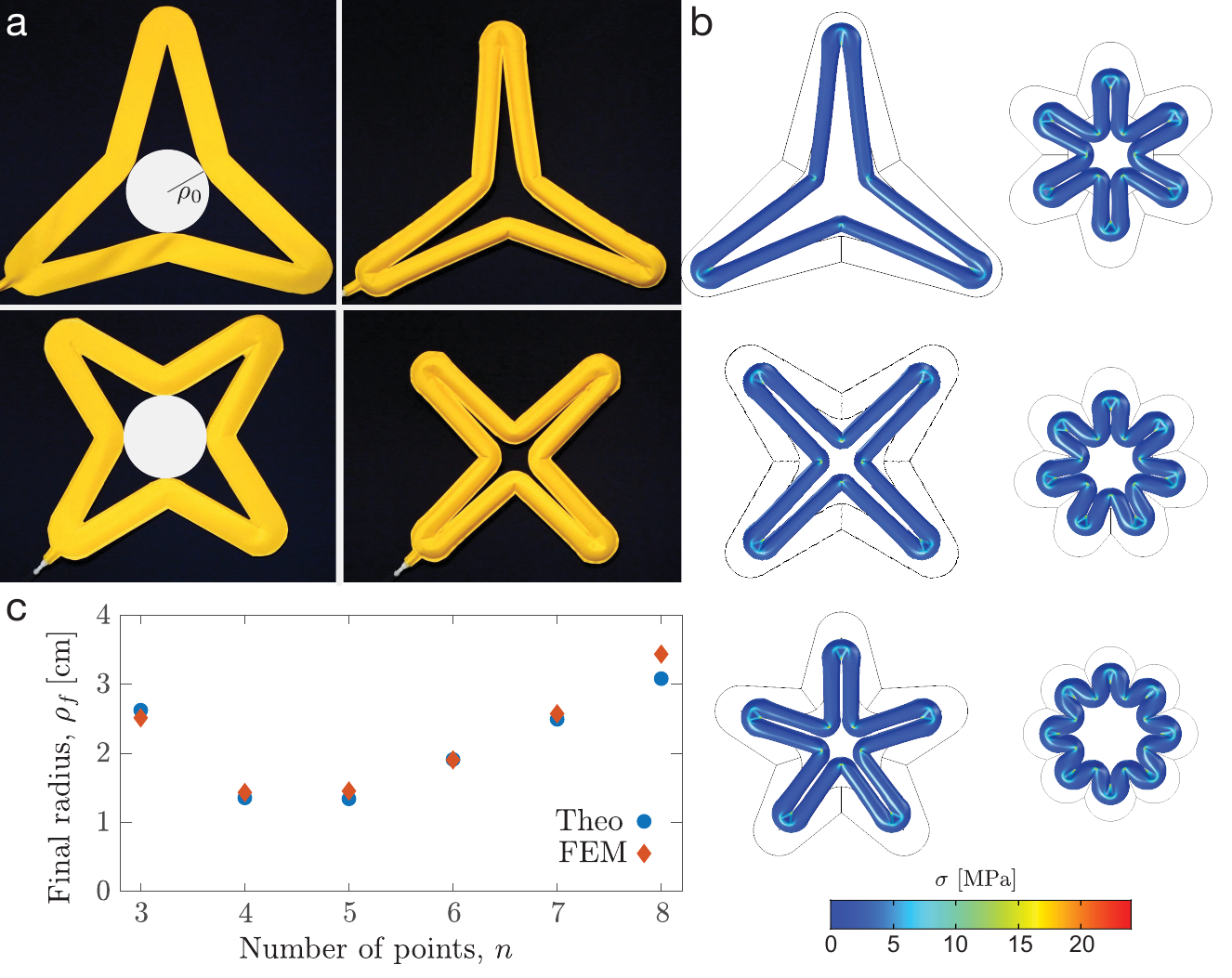}
\caption{Star geometry. (a) Experimental shapes of flat and inflated star ($n$=3, 4, $\rho_0=5$\,cm, $W=3$\,cm). (b) FEM simulations with $n$=3 to 8 with the same initial radius $\rho_0$. Von Mises stress is colorcoded. (c) Theoretical (blue dots) and FEM (red diamonds) predictions of the final inner radius $\rho_f$ as a function of $n$.}
\label{fig:star_fem}
\end{figure*}

\subsection{Optimization of the shape of the star.}  To optimize the actuation of the soft gripper, we seek the optimal shape that maximizes the star contraction $\rho_{eq}-\rho_0$, where $\rho_0=\rho(\phi_{i0})$ is the initial inner radius. As a design guideline, we choose internal angles, $\phi_{i0}=30^\circ$, that guarantee a complete closure of a single actuator. In the ideal zero-width limit and large $n$, the inner radius of the star should reduce to zero upon inflation. In fact, finite width effects and interaction between internal and external stiffnesses decrease the maximum contraction of the star. Figure\,\ref{fig:star_fem}a shows the target shape of stars with 3, 4 and 5 points with the same initial inner radius $\rho_0 =5$ cm. % (Movie S2). 
In Fig.\,\ref{fig:star_fem}b, we numerically compute the target shape of the gripper by varying the number of points of the star ($n$=3 to 8) for the same initial inner radius. Notice that $\ell$ decreases with $n$, until eventually it becomes negative for $n=9$. The radial contraction is maximal for $n$=4 or 5 (Fig.\,\ref{fig:star_fem}b) reaching 74\% of contraction. The final internal radius $\rho_f$ can also be predicted analytically as a function of $n$ (blue dots in Fig.\,\ref{fig:star_fem}c) by taking into account the finite width of the tubes in good agreement with the numerical calculations (Star Geometry with Finite Width, Supplementary Material).

\subsection{Operating pressure}
We now discuss the effect of the actuator size on the operating pressure and the force developed by the star gripper. The actuation kinematic is independent of the star scale (Fig. S5, Supplementary Material for a demonstration of a large-scale gripper). If the dimensions of the star are amplified by a factor $\eta$, $(\rho_0, R, W, \ell) \rightarrow (\eta \rho_0, \eta R, \eta W, \eta \ell)$, but the thickness $t$ is kept the unchanged, the pressure required to obtain an equivalent contraction scales as $p \rightarrow p/\eta^3$ and the gripper stiffness as $K\rightarrow K/\eta^2$. If the thickness of the sheet is also scaled as $t\rightarrow \eta t$, the operating pressure range does not change and the new stiffness scales as $K\rightarrow \eta K$.

A limitation of the design is the failure of the gripper observed at very high pressures for which the seal tends to fail at the inner edge of the hinge section. If $T_{max}$ is the maximum vertical tension the edge can withstand, then the system fails at a maximum pressure $p_{max}\sim T_{max}/W$\,\cite{siefert2019inflating}. If the dimensions are scaled by $\eta$, then the maximum supported pressure must be scaled as $p_{max} \rightarrow p_{max}/\eta$, and then $K \rightarrow K$.

\section{Performance of the fabric-based star soft robotic gripper}

\begin{figure*}%[tbhp]
\centering
\includegraphics[width=16cm]{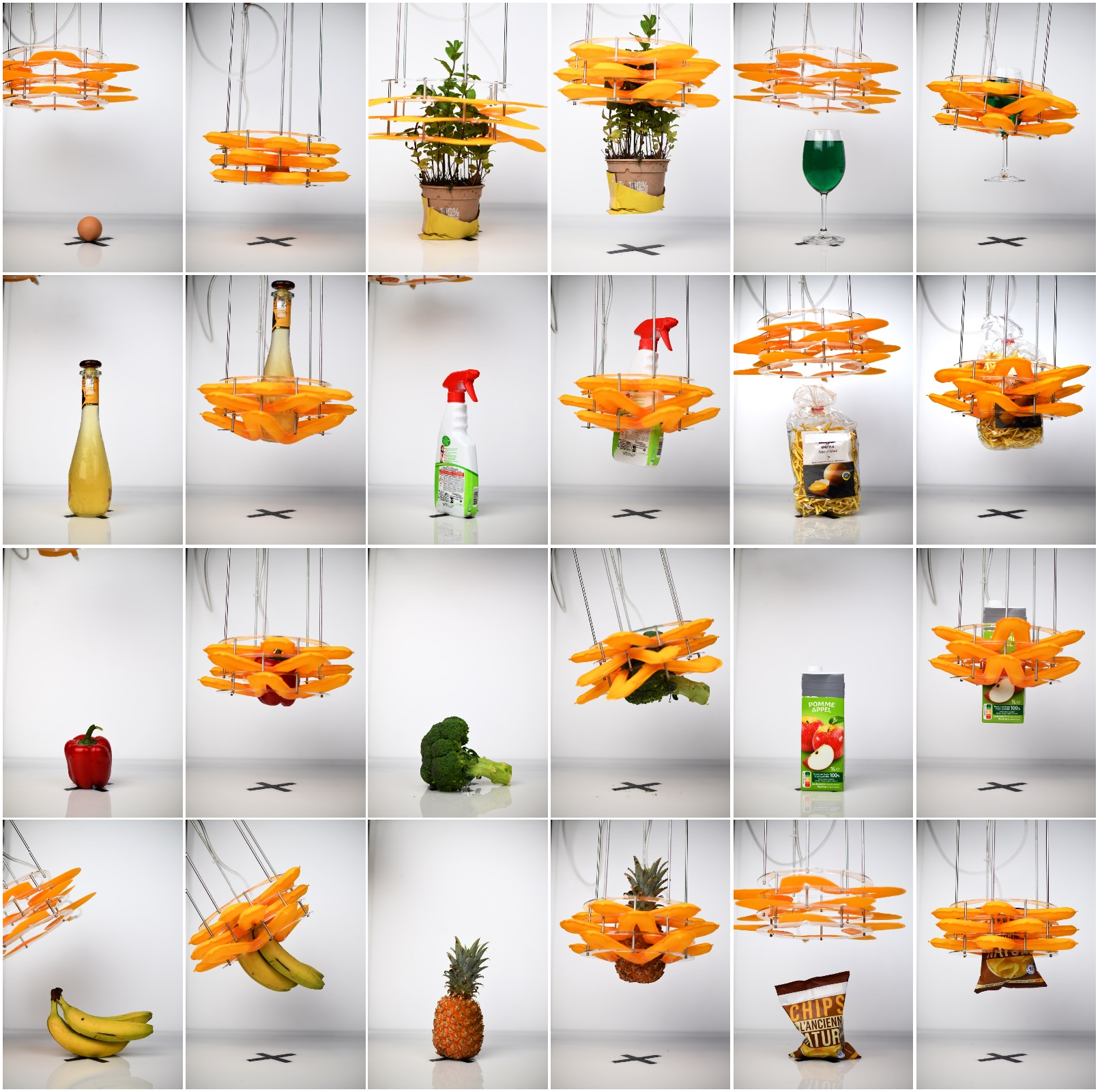}
\caption{Versatility of the gripper. We demonstrate that the gripper can move everyday objects of different shapes, weights and rigidities: an egg, a mint plant, a wine glass, an alcohol bottle, a household product bottle, a bag of dry pasta, a bell pepper, a broccoli head, a juice box, a bunch of bananas, a pineapple and a bag of chips.}
\label{fig:manyshapes}
\end{figure*}

The robustness of the manufacturing process allows us to easily produce multiple stars to further increase grip strength by stacking multiple stars. When $\rho_0$ is large or the object is too heavy, out-of-plane deformations can be observed while the arms remain straight (see Fig.~\ref{fig:star}d where the inner corners of the star are pushed down). The addition of a rigid frame solves this problem by constraining the arms to the plane. This frame also allows to position the gripper around the object when not pressurized by keeping the flat sheets in the plane.

We increase gripping efficiency by coating the inner corners of the star with a thin silicone-based elastomeric film (Fig.\,\ref{fig:star}e, and Coating Process, Supplementary Material). We coat the fabric with a polymer solution that forms a thin, almost uniform layer as the film cures\,\cite{lee2016fabrication} to improve friction and object grip. 

We test this design for a stack of three stars enclosed in a rigid frame consisting of four acrylic frames connected by stainless steel optical posts (ThorLabs mini-series). We first demonstrate the ability of the gripper to grasp a variety of objects ranging from delicate objects such as a garden mint to complex shaped objects such as eggs, flat bottles, broccoli heads and a bunch of bananas with the same gripper (Fig.\,\ref{fig:manyshapes}). We then measure the maximum weight the gripper can hold to lift a cylindrical container for an arbitrary pressure $p=37\,$kPa. While the uncoated fabric can only hold a mass of $2.2\,$kg, the coated model manages to move masses greater than $8.7\,$kg. Note that the mass of the three actuators amounts to 60 g, making the lifting ratio (object mass/clamp mass) greater than 100, comparable to the best performance observed for fluid elastomeric actuators and using electro-adhesion\,\cite{shintake2018soft}.

The gripper can lift heavy objects because we take advantage of the high bending stiffness of the inflated tubes. The bending stiffness (per unit width) of a thin-walled inflated cylinder $EW^2 t/(2 \pi)$ is indeed much higher than $B$\,\cite{main1995beam}. 

Other advantages of our gripper over other strategies are the simplicity of manufacturing compared to elastomeric gripping methods that require multiple molding steps. The stars can be printed directly in the plane unlike devices that require additional folding or pleating\,\cite{nishioka2017development, cappello2018exploiting, niiyama2015pouch}. 

The actuation strategy is scalable due to the predominance of geometry in the tube actuation mechanism. The gripper is also easily transportable and foldable when not under pressure, like other fabric-based designs. 

Finally, we consider the limitations of our gripper. First, the size of the object the gripper can grasp is limited by the two diameters of the gripper before and after inflation (smaller than $\rho_0$ and larger than $\rho_f$). Although the gripper is very versatile for gripping objects of various geometries, it is not effective for gripping very thin objects such as plates or thin disks.

We envision that our approach of using the actuation of fabric-based inflatable structures, will open new possibilities in developing soft robotic matter on a larger scale capable, of performing complex tasks after the application of a simple stimulus.

%\appendix

\section{Experimental Section}
\subsection*{Fabrication of actuators}
The inflatables are made by sealing together two thermo-sealable nylon fabric sheets (impregnated with thermoplastic urethane) using a temperature-controlled soldering iron unit (PU81 from Weller) mounted on a CNC machine (Aureus 3X 10 from Euromakers). The temperature was set at $200^\circ$\,C and the travel speed was set at $0.1$ mm/s to ensure a good seal. The 2D designs were drawn in Fusion 360 CAD software  and then exported to G-code. The fabric used has a thickness of $0.3$\,mm and a measured Young's modulus of $E = 128$\,MPa. We coat the fabric with vinylpolisiloxane (VPS Elite Double 32). The typical thickness of the elastomeric coating was 200\,$\mu m$ and the Young's modulus of the elastomer was $E_f=1.1$\,MPa (Coating Process, Supplementary Material).

\subsection*{Mechanical characterization}
The prototypes were inflated by connecting them to an air compressor and the pressure was controlled by a pressure regulating valve. All forces were measured in a universal testing machine (Zwick Z2.5TH) with XForce HP 10 N load cells. 

\subsection*{Numerical Methods}
All finite element simulations were performed in COMSOL Multiphysics 5.6 using the Shell interface included in the Structural Mechanics Module. All the simulations were carried out with a linear elastic Hookean material model with geometric nonlinearities. We searched for solutions with the default stationary solver, where the non-linear Newton method has been implemented. Mesh refinement studies were undertaken to ensure convergence of the results. Details on the modelling of our inflatables are found in Finite Element Simulations, Supplementary Material.

\begin{acknowledgments}
We thank Victor Charpentier for discussions on the design and practical applications of the gripper. This work was supported by CNRS-Momentum and ANR BioSoftAct (ANR-22-ERCS-0006). 
\end{acknowledgments}

\bibliography{bib_v2}% Produces the bibliography via BibTeX.

\pagebreak
\widetext

\setcounter{figure}{0}
\setcounter{equation}{0}

\appendix

\section*{Supplementary Material}

\section*{Inflation of an annular sector}
\subsection*{Maximization of the volume of a ring sector}
\begin{figure}
\centering
\includegraphics[width=11.8cm]{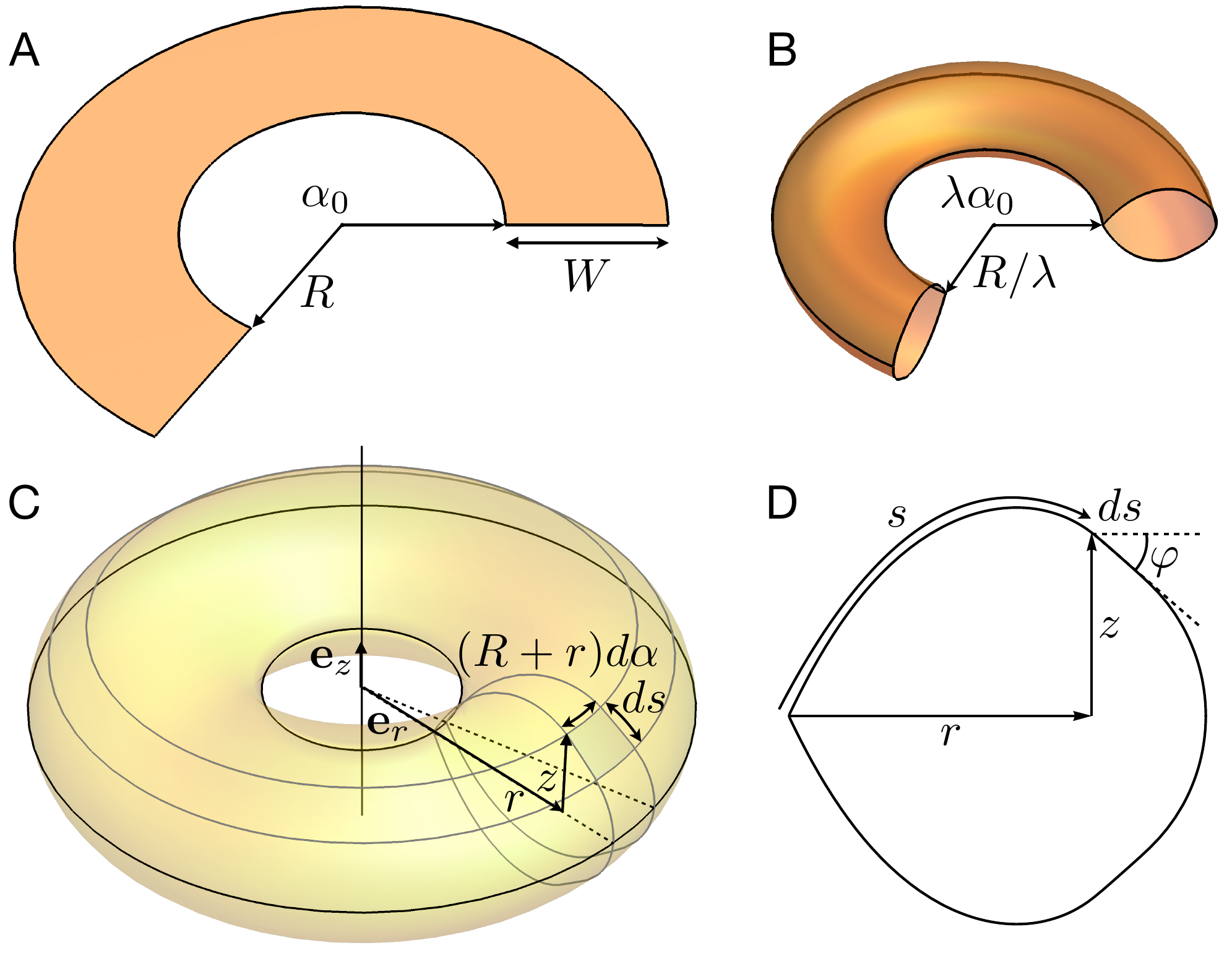}
\caption{(A) Curved section of a V-actuator with inner radius $R$, width $W$ and arc $\alpha_0$. (B) FEM simulations of the actuator showing overcurvature during inflation. (C) Schematic of a closed ring showing coordinates used. (D) Cross section of the ring predicted by the theory.}\label{fig:overcurvature-profile}
\end{figure}

Figure~\ref{fig:overcurvature-profile}A shows a sector of a ring representing the curved section of a single V-actuator with inner radius $R$, width $W$ and initial angle $\phi_0 = \pi-\alpha_0$, where $\alpha_0$ is the arc of the ring in the reference (flat) configuration. After inflation, the new radius of curvature of the inner edge is $R/\lambda$ and the new arc is $\alpha=\lambda \alpha_0$, where $\lambda$ is the coiling factor (see Figure~\ref{fig:overcurvature-profile}B for the FEM solutions showing overcurvature after inflation). Recall that the actual angle $\phi$ of the actuator and the coiling factor are related by:
\begin{equation}
    \lambda = \frac{\pi-\phi}{\pi-\phi_0}.
\end{equation}
The shape of the inflated structure is described by the vector~\cite{siefert2019programming}:
\begin{equation}
    \mathbf{r}(s,\theta) = (R/\lambda+r(s)) \mathbf{e}_r (\theta) + z(s) \mathbf{e}_z
\end{equation}
where $\mathbf{e}_r (\theta)$ and $\mathbf{e}_z$ are, respectively, the radial and vertical cylindrical vectors, $\theta$ is the polar angle, $s$ is the radial coordinate along a meridian, $z(s)$ the vertical coordinate and  $r(s)$ is the radial distance measured from the inner edge (see Figure~\ref{fig:overcurvature-profile}C).

Upon inflation, the system minimizes the total energy $U_T = U_{s} - pV$, where $U_{s}$ is the strain energy, $p$ the pressure and $V$ the total volume. In the case of an infinitely thin membrane, the total energy reduces to $-pV$, the minimization of which is equivalent to the maximization of the volume. 
For a given profile of the cross section, the volume of the sector of the ring is given by:
\begin{equation}
    V = 2 \lambda \alpha_0 W^3\, \int^1_0 \left( \frac{R^*}{\lambda} + r^*(s^*) \right) z^*(s^*) \cos{\varphi(s^*)} \, ds^*, \label{eq:vol}
\end{equation}
where the notation $()^*$ indicates the nondimensionalization of lengths by $W$.

\subsection*{Inextensibility condition}
The inextensibility condition along a meridian is enforced by imposing that $s^*$ is the arc-length of the meridian profile. In the azimuthal direction, the local projected perimeter of the structure at the curvilinear coordinate $s^*$ is $\mathcal{P}_{\lambda} (s) = \alpha_0  \left[R + \lambda r (s) \right]$. Due to inextensibility condition, this perimeter must be less than or equal to the reference perimeter of the flat configuration $\mathcal{P}_{\lambda} (s) \le \mathcal{P} (s) = \alpha_0 (R+s)$.

We define $u(s)$, a parameter that quantifies the difference between the local projected perimeter and the reference perimeter:
\begin{equation}\label{eq:u}
    u^*(s^*) = \frac{\mathcal{P}_{\lambda} (s) - \mathcal{P} (s)}{\alpha_0 W} = \lambda r^* (s^*)-s^*.
\end{equation}
The inextensibility condition imposes $u^*(s^*)>0$ for any $s^*$. To describe the equilibrium shape of the inflated ring, the following augmented functional is proposed~\cite{siefert2020geometry}
\begin{equation}
    \mathcal{L} = 2 \lambda \alpha_0 \int^1_0 \left[\frac{R^*+s^*+u^*}{\lambda} z^* \cos{\varphi} + A \left(\cos{\varphi}-\frac{1+{u^*}'}{\lambda}\right) + B\left(\sin{\varphi}-{z^*}'\right) - e^{\beta {u^*}} \right] ds^*.
\end{equation}
The variables $A(s)$ and $B(s)$ are two Lagrangian multipliers that enforce the inextensibility along a meridian. To enforce inextensibility in the azimuthal direction, the term $-e^\beta u^*$ is added to penalize any stretching of the projected perimeter, where $\beta$ is a large numerical factor.

The maximization of this functional gives:
\begin{align}\label{eq:variation}
    & \left[z^* \cos{\varphi} - \beta \lambda e^{\beta {u^*}} + A'\right]\delta u^* - A\delta u^* \vert^1_0 + \left[ \frac{R^* + s^* + u^*}{\lambda}\cos{\varphi} + B' \right]\delta z^* - B\delta z^*\vert^1_0 + \nonumber \\
    &\left(\cos{\varphi} - \frac{1 + {u^*}'}{\lambda} \right)\delta A + \left(\sin{\varphi}- {z^*}'\right)\delta B + \left[-\frac{R^*+s^*+u^*}{\lambda}z^* \sin{\varphi} - A\sin{\varphi} + B\cos{\varphi} \right]\delta \varphi
    =0.
\end{align}
leading to the following system of equations:
\begin{align}\label{eq:ODEs}
    &A' = -z^* \cos{\varphi} +\lambda \beta e^{\beta u^*} \nonumber \\
    &B' = -\frac{R^* + s^* + u^*}{\lambda} \cos{\varphi} \nonumber \\
    &{u^*}' = \lambda \cos{\varphi} - 1 \nonumber \\
    &{z^*}' = \sin{\varphi} \nonumber \\
    &\left[A + z^* \frac{R^* + s^* + u^*}{\lambda} \right] \sin{\varphi} = B\cos{\varphi}.
\end{align}
Since $u^*(0)=0$ ($r^*(0)=0$) and $u^*(1)$ is free, the first boundary term in Eq.\,(\ref{eq:variation}) implies that $A(1)=0$. Moreover, $z^*(0)=z^*(1)=0$ since the inner and outer edge of the membrane are constrained in the $xy$-plane. We thus obtain the following set of boundary conditions:
\begin{align}
    A(1) &= 0  \nonumber\\
    u^*(0) &= 0 \nonumber\\
    z^*(0) &= 0 \nonumber\\
    z^*(1) &= 0.
\end{align}
This differential-algebraic system of equations can be converted into a set of first order differential equations by deriving once the last equation of Eq.\,(\ref{eq:ODEs}) and writing it as:
\begin{equation}
    \varphi' = - \frac{\sin{\varphi}}{B} \left(
    \frac{R^* + s^* + u^*}{\lambda} +\lambda \beta e^{\lambda u^*} \sin{\varphi} \right).
\end{equation}
Evaluating the last equation of Eq.\,(\ref{eq:ODEs}) at $s^*=1$, we obtain a fifth boundary condition:
\begin{equation}
    \varphi(1) = -\frac{\pi}{2}.
\end{equation}
The present boundary-valued problem is solved in Matlab using the bv4pc function (see a solution for the cross-section in Fig.\,\ref{fig:overcurvature-profile}D).

\subsection*{Target angle of an actuator} For given values of $\lambda$ and $R^*$, an initial solution with $\beta=0$ is found starting from a suitable initial guess solution. Then, new solutions are computed for increasing values of $\beta$ up to large values (>1000), ensuring the convergence of the solutions. From there, the value of $\lambda$ is varied (from 1 to 1.5) and the volume is computed using Eq.\,(\ref{eq:vol}). For a fixed value of $R^*$, there exists an optimal value $\lambda_t$ that maximizes the volume, which corresponds to the rest angle of the actuator. 

\begin{figure}
\centering
\includegraphics[width=1\textwidth]{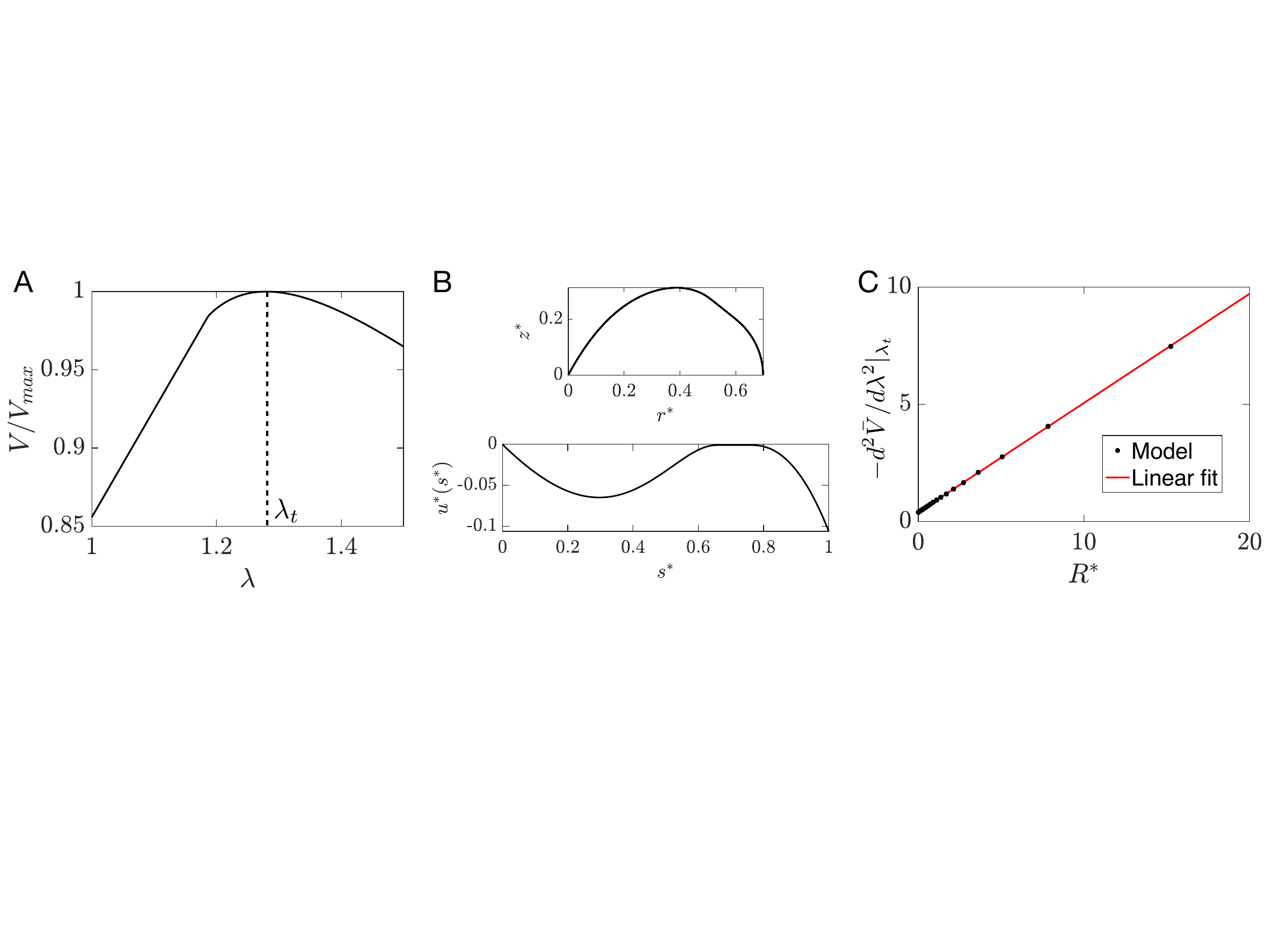}%V_vs_lamb.pdf
\caption{(A) Variation of the volume of the ring sector $V$ normalized by the maximum volume $V_{max}$ as a function of $\lambda$ for $R^*=0.1$. The optimal coiling is obtained with $\lambda_t=1.28$ which defines the target angle. (B) Profile $\{r^*(s^*),z^*(s^*)\}$ of the cross section of the optimal solution for $R^*=0.1$. Below, the function $u^*(s^*)$ is shown. (C) The term -$d^2 \bar{V}/d\lambda^2 \vert_{\lambda_t}$ is plotted as a function of $R^*$.}\label{fig:V-vs-lamb}
\end{figure}

Fig.~\ref{fig:V-vs-lamb}A shows the variation of the volume $V$ as a function of the coiling factor $\lambda$ for $R^*=0.1$. The optimal coiling maximizing the volume is obtained with $\lambda_t = 1.28$. The cross sectional profile of the optimal solution for $R^*=0.1$ and parameter $u^*(s^*)$ are plotted in Fig.~\ref{fig:V-vs-lamb}B.

\subsection*{Discontinuity in $V'(\lambda)$} 
\begin{figure}
\centering
\includegraphics[width=.7\textwidth]{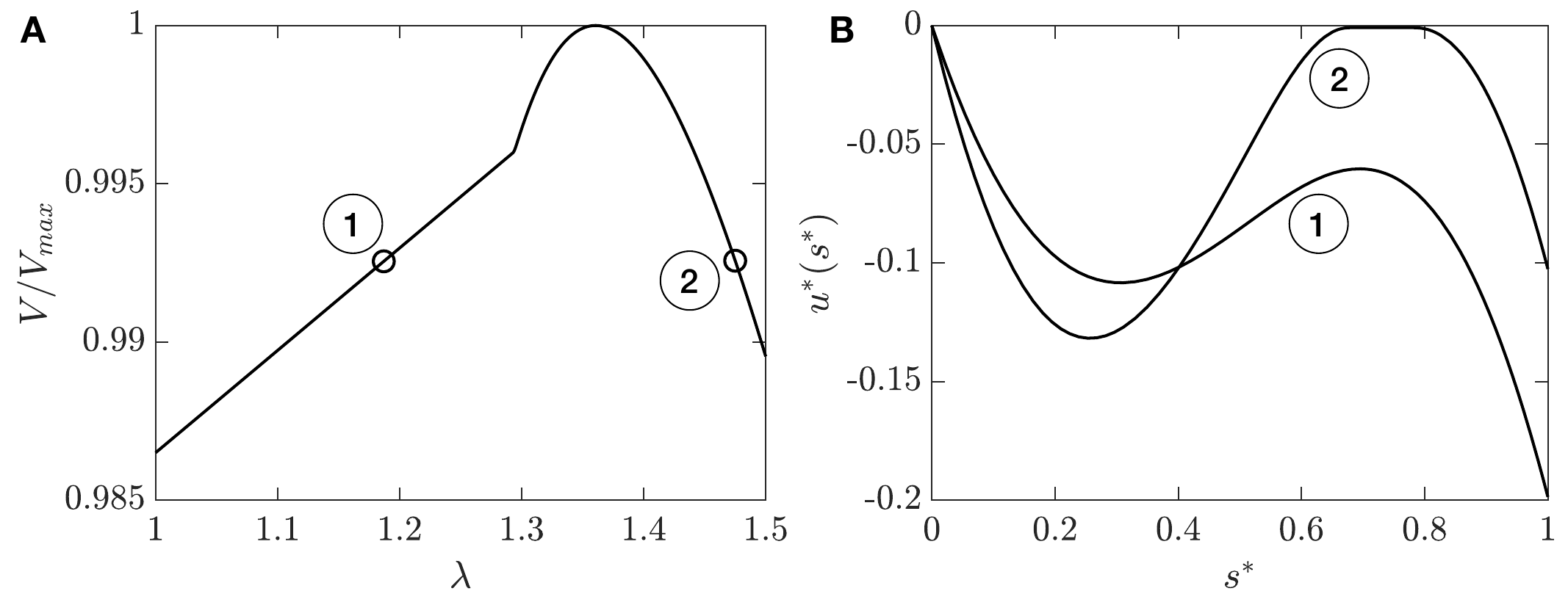}
\caption{(A) Variation of the normalized volume of a ring sector with $R^*=10$, where two solutions are labelled by  \textcircled{\raisebox{-0.9pt}{1}} ($\lambda=1.2$) and \textcircled{\raisebox{-0.9pt}{2}} ($\lambda=1.475$). (B) $u^*(s^*)$ profiles of the two labelled solutions.}\label{fig:discontinuity}
\end{figure}

The numerical computations of the volume showed a sharp discontinuity in $V'(\lambda)$, which results in a abrupt jump of the moment $M$ that we do not observe neither in the experiments nor in FEM.
This situation occurs when opening the V-actuator and corresponds to the physical limit where there are no regions in azimuthal tension in the structure rather than in the inner rim ($u^*(s^*)<0$ for all $s^*>0$).
Fig.~\ref{fig:discontinuity} illustrates this situation for $R^*=10$, where the discontinuity is more pronounced. 

\section*{Finite width effect.}
\subsection*{Star geometry with finite width} Fig.~\ref{fig:star_sector_diagram} shows a schematic of a sector of a $n$-pointed star of width $W$ and inner radius $\rho_0$. Note that the inner and outer corners have been designed with radius of curvature zero. If $\rho_0$, $W$ and $\phi_{i0}$ are chosen, then length of the arms is given by
\begin{equation}
    \ell = \frac{(\rho_0+W) \sin{\pi/n}-W \cos{\phi_{i0}/2}}{\sin{\phi_{i0}/2}}. \label{eq:l}
\end{equation}
The above equation shows that $\ell$ decreases with the number of points and eventually becomes negative.

We performed FEM analysis of the star gripper upon inflation. We simulate a only sector of the star using a sheet with the shape shown in Fig.~\ref{fig:star_sector_diagram} and applying $n$-fold cyclic symmetry. 
The final radius after inflation can be estimated by clearing $\rho_0$ from Eq.\,(\ref{eq:l}) and replacing $W$ by $W'=2W/\pi$ (we assume that the new width at the corners and at the arms are similar), and $\phi_i$ by $\phi_{if}=\max(0,\phi^{eq}_i)$. The final radius of the star is given by
\begin{equation}
    \rho_f = \frac{\ell \sin{\phi_{if}/2} + W'(\cos{\phi_{if}/2}-\sin{\pi/n})}{\sin{\pi/n}}.
\end{equation}

\subsection*{Gripper strength} When the width of the tube is considered, the actual inner radius of a star with inner angles $\phi_i$ is
\begin{equation}
    \rho(\phi_i) = \frac{\ell \sin{\phi_i /2} + W'(\cos{\phi_i/2} - \sin{\pi/n} )}{\sin{\pi/n}}.
\end{equation}
The radial force is
\begin{equation}
    F(\rho) = \frac{dE}{d\rho} = \frac{d \phi_i}{d\rho} \frac{d E}{d\phi_i}.
\end{equation}
The radial force vanishes at an equilibrium radius $\rho_{eq} = \rho(\phi^{eq}_i)$. A small radial displacement above $\rho_{eq}$ produces a radial force that, at first order, reads
\begin{equation}
    F(\rho) = 4 n (C_i + C_e) \left( \frac{\sin{\pi/n}}{\ell \cos{\phi^{eq}_i}/2 - W' \sin{\phi^{eq}_i}}\right)^2 (\rho - \rho_{eq}).
\end{equation}

\section*{Finite element simulations}
\begin{figure}
\centering
\includegraphics[width=.9\textwidth]{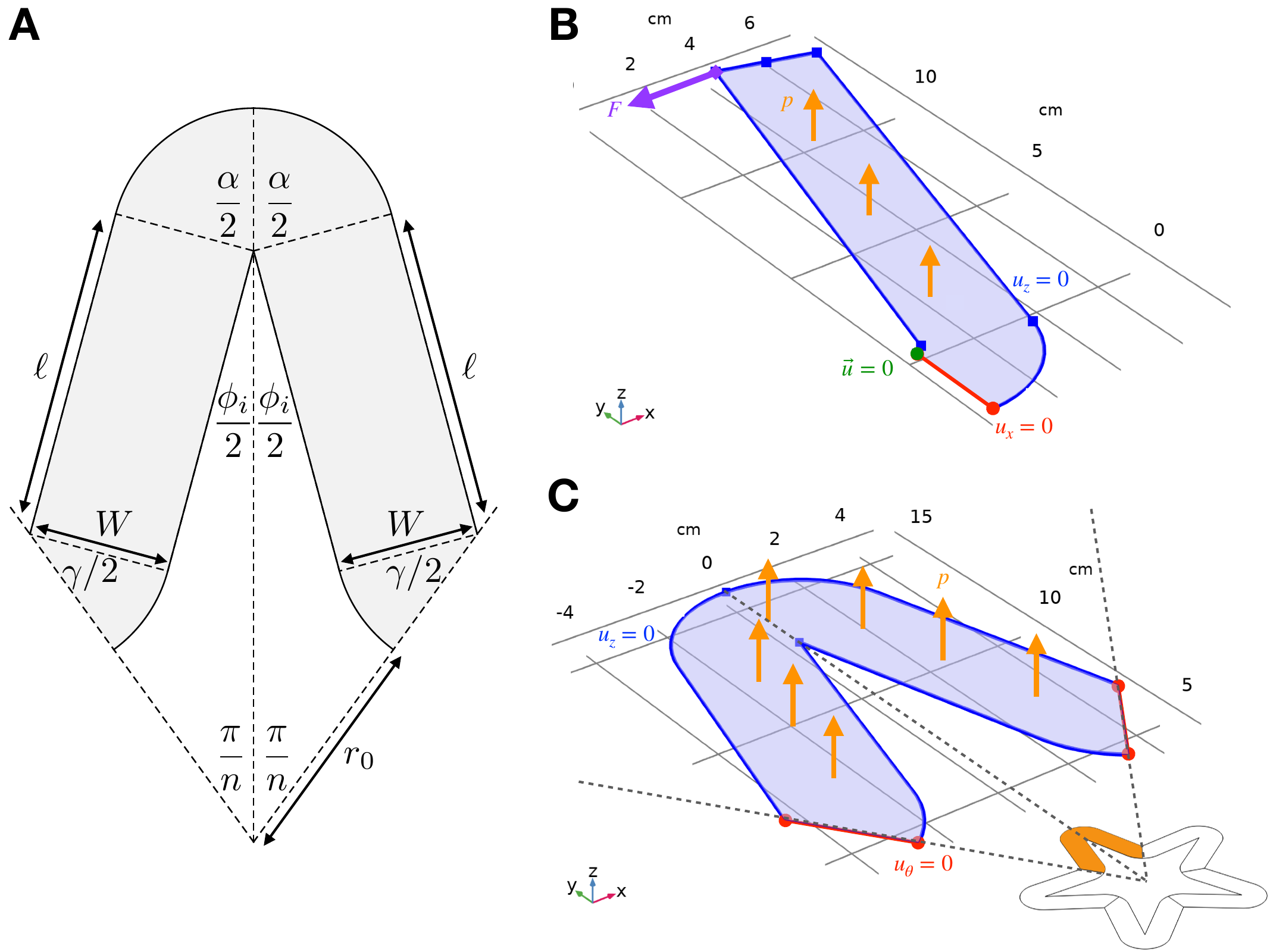}
\caption{(A) Schematic of a sector of a $n$-pointed star. Here, $\alpha = \pi - \phi_i$ and $\gamma=\pi-2\pi/n-\phi_i$. Boundary conditions used in FEM simulations of a V-actuator (B) and a sector of an $n$-points star (C).  }\label{fig:star_sector_diagram}
\end{figure}

We model just the upper sheet of our inflatables as initially flat plates where the rims are constrained in the $xy$-plane. Then, a pressure $p$ is applied to the surface plate.
For the V-actuator (Fig.~\ref{fig:star_sector_diagram}B), we model just one half of the actuator defined its plane of symmetry. The points lying on the plane of symmetry are constrained to this plane and the interior point of the hinge is fixed.
An additional force $F$ pointing in the $x$-axis is applied at the end of the arm to probe the mechanical response of the actuator at fixed pressure.
For a $n$-pointed star, we model one circular sector of angle $2\pi/n$ (Fig.~\ref{fig:star_sector_diagram}C). In cylindrical coordinates, we prescribe zero azimuthal displacement to the points lying on the boundary planes of the sector.

\section*{Coating process.} 
Figure 6 illustrates the process of coating the inner corners of the star to increase friction and adhesion between the gripper and the objects. The star is protected by a mask. A liquid silicone-based polymer solution is poured onto the star and drains under gravity.  Over time, the cross-linking of the polymer film that emerges from the drainage process produces a thin elastic film  on the inner corner of the stars.

\begin{figure}
\centering
\includegraphics[width=.7\textwidth]{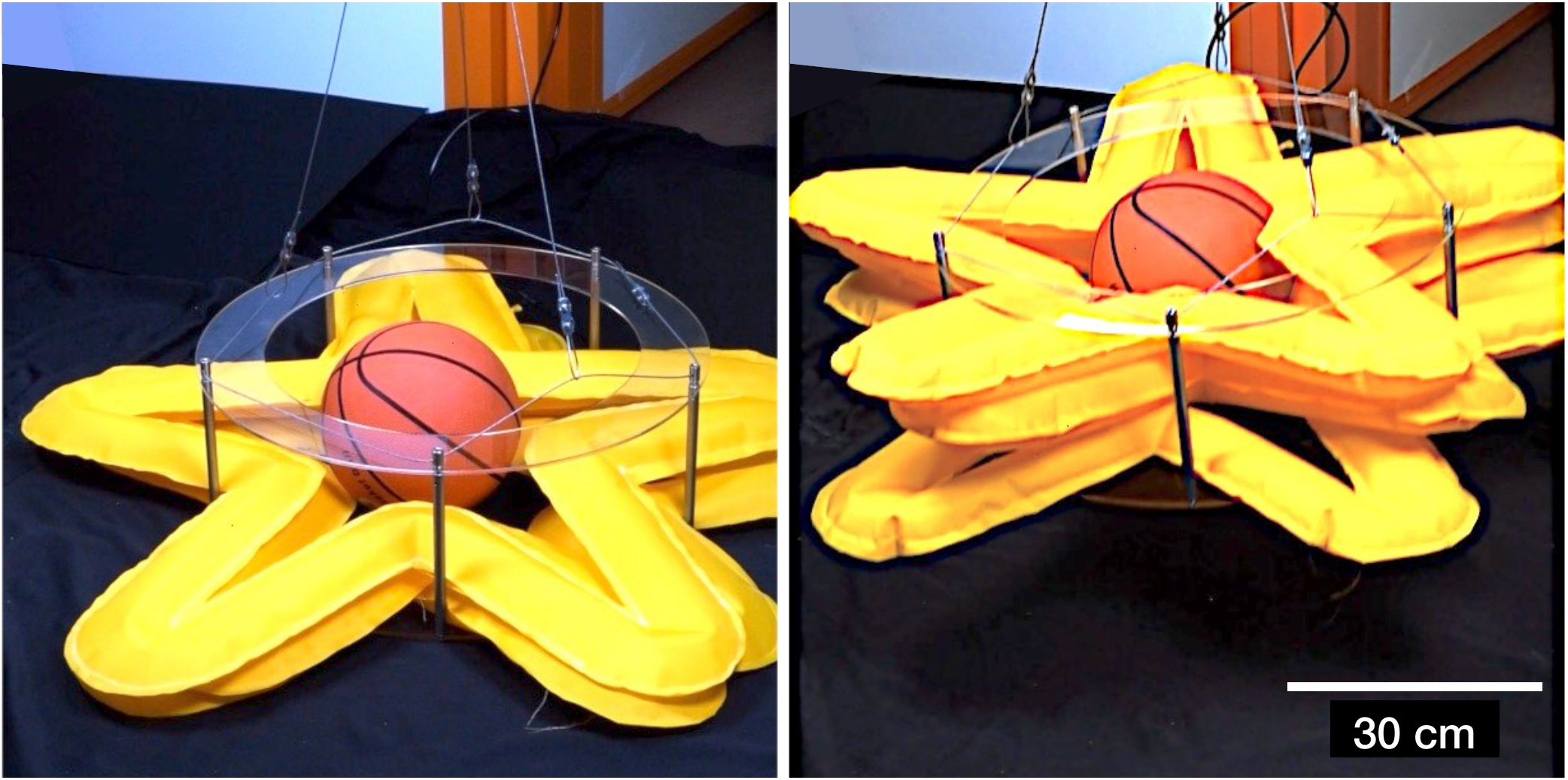}
\caption{A large star gripper gripping a basketball.}\label{fig:basketball}
\end{figure}

\begin{figure}
\centering
\includegraphics[width=.7\textwidth]{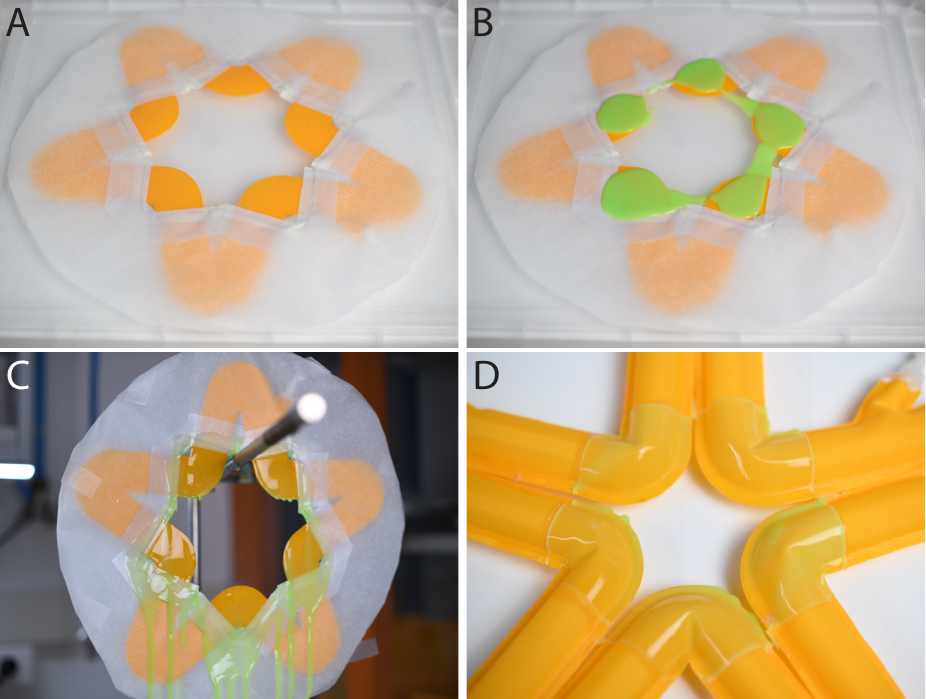}
\caption{Elastomer coating process to increase friction and adhesion with the object. (A) The fabric star is protected by a mask. (B) The molten polymer is poured onto the fabric. (C) The molten polymer flows under gravity and cures in finite time. (D) The mask is removed after curing.}\label{fig:coating}
\end{figure}

\end{document}